%% file: relatorio.tex
\documentclass[11pt,oneside,a4paper]{report}

\usepackage[english]{babel}
\usepackage[utf8]{inputenc}
\usepackage{amsmath,cite}
\usepackage{graphicx}
\usepackage{color}
\usepackage{subfigure}
\usepackage{comment}
\usepackage{url}
\usepackage{balance}
\usepackage{algorithm}
\usepackage[noend]{algpseudocode}
\inputencoding{utf8}
\usepackage{csquotes}

\begin{document}

\title{%
Mixtape Application:\\
\textbf{\Large{Music map methodology and evaluation}}}
\author
{\bfseries{Pedro H. F. Holanda \hspace{1cm} Bruno Guilherme \hspace{1cm} Luciana Fujii} \\
  \bfseries {Ana Paula C. da Silva \hspace{1cm} Olga Goussevskaia}\\\\
  Computer Science Department\\ Universidade Federal de Minas Gerais (UFMG), Brazil\\\\
{\tt\small holanda@dcc.ufmg.br, brunoguilherme@dcc.ufmg.br, lucianafujii@dcc.ufmg.br}\\
{\tt\small ana.coutosilva@dcc.ufmg.br, olga@dcc.ufmg.br}
}
\date{May 25, 2016}
\maketitle

\tableofcontents

\input{sections/introduction}

\input{sections/methodology}
\input{sections/tvDomain}
\input{sections/musicDomain}
\input{sections/conclusion}

\newpage

\inputencoding{latin1}
\bibliographystyle{abbrv}
\bibliography{relatorio}  
\inputencoding{utf8}

\end{document}

%% file: sections/introduction.tex
\chapter{Introduction}\label{sec:introduction}

In the old days, people used to store their media collections (books, music
records, video tapes, etc) in dedicated cabinets or shelves, sometimes as part
of the home decoration. The organization would be done alphabetically or chronologically.
With the advent of digital media and compressed file formats, 
media items have become digital files, stored in extensive hierarchies of
directories on personal computers and, later, on portable mobile devices,
such as e-readers and media players. Nowadays, the portable player has fused
with the smartphone and media collections have been migrating into the cloud.

Internet storage and streaming services, like digital libraries, Youtube,
Instagram, Spotify and Google Music, have become the state-of-the-art in terms
of storage and access to media collections. They allow users to upload their
content into the cloud and browse it from any device or, for a monthly fee, have
unlimited access to the entire cloud. 

To explore the vast content of ever growing personal collections and of the
cloud, there is a necessity of developing novel ways to navigate these media
files.
Navigation functions typically available are based on filtering by attributes,
like title, artist or genre, or return a list of similar items,
computed using collaborative filtering techniques.
Some systems, after a user has used them for long enough,
are able to make taste-based recommendations based on usage
patterns.
Other systems are based on curated lists of newly released and trending
items or artists, such as the user-made playlists on
mixcloud\footnote{\url{www.mixcloud.com}} or the professional dj-curated
streams on 22tracks\footnote{\url{22tracks.com}}. 

Our approach to organize these media items is to look at how people behave on
the internet - more specifically on Online Social Networks (OSNs).
These websites are meant to extend the individual experience of society by bringing social
activities to the internet, allowing people to discuss their favorite subjects and share
their thoughts, opinions and feelings in a distributed and independent way, making them
an interesting source of many different kinds of information.

We define media item-to-item similarity based on user-generated data, 
assuming that two items are similar if they frequently co-occur in a
user's profile history. For example, if two books are frequently read
by the same people, they must have something in common.

We look specifically into two OSNs: Last.Fm and tvtag. Last.fm is an online
social network for music fans that collects the songs listened by their users
automatically. tvtag is a online social network for TV usage, where users
check-in to the TV programs and movies they watched.
For Last.fm we collected users' top 25 and top 100 most listened to songs for
approximately 380K users, while for tvtag we collected the users' entire
check-in history for 1.2M users.

Collected co-occurrence data is usually sparse (not all pairs of items will have
co-occurred at least once in the collected dataset) and nevertheless might occupy
a lot of memory space ($\Omega(n^2)$, where $n$ is the size of the collection).
To guarantee $O(n)$ space complexity and $O(1)$ query complexity of all-pairs
similarity information, we transform the collected pairwise co-occurrence values
into a multi-dimensional Euclidean space, by using nonlinear dimensionality
reduction \cite{TenenbaumEtAl2000}.

From this Euclidean space it's possible to retrieve an item's most similar item
in constant time. This allows the implementation of efficient navigation methods
which are adequate for large collections and usage on devices with small
computational power, such as mobile devices.

In this document we will focus on the similarity computation and dimensionality
reduction of the co-occurrence data, evaluating the resulting Euclidean spaces
created.


This paper is organized as follows. We discuss the used methodology in 
Section~\ref{sec:methodology}. In Section~\ref{sec:tvDomain} we evaluate an
embedding created for TV Domain based on tvtag data.
In Section~\ref{sec:musicDomain} we show maps created for the music domain based
on Last.fm data collection.
Finally, in Section~\ref{sec:conclusion}, we present our conclusions.

%% file: sections/methodology.tex
\chapter{Methodology} \label{sec:methodology}

\section{Cosine similarity}

In order to obtain similarity values between items in our database, we use
collaborative filtering techniques. These techniques are based on the fact that
items frequently co-occur in usage data, establishing similarity measures from
that data. They have proven to be successful in many situations, and have been
implemented in many renowned online systems, such as the Amazon recommendation
engine.

Just as Amazon uses the fact that two items were bought by the same person to establish a relationship between them, our studies assume that two songs are related if they are frequently played by the same user. The same assumption will be used to analyze movies and series.

We use two social networks as data sources: Last.FM and tvtag. Last.FM is a music habit sharing network. The site offers the option of installing a ``scrobbler'', which is a small software listener that automatically logs the musical activity it detects (Spotify or iTunes, for instance). tvtag, on the other hand, requires that users voluntarily register their habits, by posting comments, ``liking'' or ``checking in'' to shows on their website profiles.

Simply counting the number of individual occurrences of a given pair of
items to calculate pairwise similarity overestimates the similarity of
popular items, which will have an elevated number of individual
occurrences. That makes them clearly have a high probability of being
associated to a user. To overcome that issue, it is necessary to
transform the individual occurrence counts in a uniform similarity
measure that represents the similarity between two given items regardless
of their individual popularity. We chose to use Cosine coefficient.

Cosine is a measurement of the similarity between two vectors, expressing
the cosine of the angle between them.
The Cosine is a measurement of
orientation, not magnitude: Two vectors with the same orientation have a
cosine similarity of one, whereas two vectors 90 degrees apart have a
similarity of -1, regardless of their magnitude.

\begin{equation} cos(A,B) = \frac{cooc(A,
B)}{\sqrt{|A||B|}},\label{eq:cosine} \end{equation}
\noindent where $|A|$ and $|B|$ are the number of individual occurrences of A
and B and cooc(A,B) are the number of co-occurrences between them.

\section{Embedding techniques}

Among the dimensionality reduction we
decided to initially use the Isomap \cite{TenenbaumEtAl2000} since it has good results in many problems, such as \cite{silva2002global,van2009dimensionality,yan2007graph}. Also because one of its main properties is that, being a global dimensionality reduction technique, focuses on keeping near the points that were close in the initial space, and distant points that were far away. This property is very important for our problem, it is necessary that similar songs stay close, and different songs far away in the generated space.

\subsection{Isomap}

Isomap, first proposed by \cite{TenenbaumEtAl2000}, is a global nonlinear dimensionality reduction method based on MDS \cite{cox}, and follow these three steps.

\textbf{Distances graph:} The first part is to construct the neighborhood distances item-to-item graph. To build this graph we define the vertices as each track and the weight of an edge between track $i$ and track $j$ as $w(i,j) = 1 - cos(i,j)$ with a time complexity $O(ne)$, we define $n$ as the total number of vertices and $e$ the total number of links between them.

\textbf{Geodesic distances matrix:} Next we compute the shortest path between all vertices, building a $n x n$ matrix, the time complexity of this step is $O(ne + n^2 log n)$.

\textbf{Low-dimensional embedding:} In order to build a low-dimensional embedding we use the classical multidimensional scaling (MDS) algorithm with complexity $O(n^3)$.

The final result is a low-dimensional Euclidean space where each point represents a track and the distance between two tracks represent how different they are.

\subsection{L-Isomap}

L-Isomap is an approximation of Isomap that addresses the last two steps time complexity choosing $l$ landmark points, where $l << n$. With the distances graph, obtained in Step 1 of Isomap, we follow the next steps.

\textbf{Geodesic distances matrix:} Compute the shortest path between the vertices and the landmarks only, building a $l x n$ matrix, so the time complexity of this step is lower $O(le + l n log n)$.

\textbf{Low-dimensional embedding:} To build the low-dimensional embedding we use the Landmark MDS (LMDS) algorithm, where we find the Euclidean embedding in $O(l n^2)$ time~\cite{silva2002global}.

\textbf{Distance-based triangulation:} To calculate the $l - n$ remaining points we use the low-dimensional space obtained in the previous step to triangulate these points as Step 3 in \cite{de2004sparse}.

With this variant of Isomap we are able to build bigger maps in a lower amount of time, but with a decrease in the accuracy which will be analysed in Section~\ref{sec:musicDomain}.

To select the landmark points we used two different ways that will be analysed in Section~\ref{sec:musicDomain}:

\begin{itemize}
  \item Random choice.
  \item MaxMin function: suggested by \cite{de2004sparse}, $s$ seed landmarks, chosen randomly such as $s << l$, and the other landmarks are chosen one at a time, where each new landmark maximizes the minimum distance between the landmarks and the unused data points.
\end{itemize}

%% file: sections/tvDomain.tex
\chapter{TV Domain}\label{sec:tvDomain}

The method of collecting user data and constructing maps based on co-occurrence
data can be applied to different media domains, such as movies, TV shows,
books, or music.  This chapter will evaluate the application
of these methods on the TV domain based on data collected from tvtag and TMDB (The Open Movie Database).

Tvtag is an Online Social Network where users added their TV watching
history checking-in, liking or disliking watched TV shows and movies.
The data was collected between 2011 and 2012 and consists of 29M
check-ins from users in approximately 3,000 TV programs and 7,000 movies.

We use the fact that one person likes two TV shows or movies as an
indication of their similarity. The more two TV shows or movies co-occur
in users likes, the more similar they are. To compensate for the
popularity of the TV shows or movies, we used the cosine coefficient.

We reduce the dimensionality of the map using Isomap 
\cite{TenenbaumEtAl2000} and classic MDS \cite{Kruskal1978}.

\section{Gender-based evaluation}

To evaluate the quality of the generated map, we collected data from an
independent source, TMDB, that contains gender for TV shows and movies.
We categorized approximately 3000 shows with 26 different genders, with
the majority of the shows belonging to more than one gender.

We expect items that are close in the map (similar items) to be of the
same or similar gender. Since most items belong to more than one gender,
it would not be enough to simply check if two items belong to the same
gender, therefore we defined gender similarity using the co-occurrence of
two genders in the users' like history.

The similarity of two genders $g_i$ and $g_j$ is defined as:
\begin{equation} simBetweenGenders(g_i,g_j) = cos(g_i, g_j) = \frac{cooc(g_i,
g_j)}{\sqrt{items(g_i) items(g_j)}},\label{eq:simPorGen} \end{equation}
\noindent where $cooc(g_i, g_j)$ is the number of items that were
classified with both genders and $items(g_i)$ and $items(g_j)$ are the
number of items that were classified with at least gender $g_i$ or at
least gender $g_j$, respectively.

As in our database most items belong to more than one gender, we define
the gender-based similarity between two items A and B as the average of the cosines between the genders of item A and item B:
\begin{equation}
genderBasedSimilarity(A, B) = \frac{\sum_{g_i \in GA, g_j \in GB}{simBetweenGenders(g_i,
g_j)}}{kl}.\label{eq:simGenero}
\end{equation}
\noindent where 
$G_A = <g_1, g_2, \cdots g_k>$ and $G_B =<g_1, g_2, \cdots g_l>$

\textbf{Gender gradient analysis in the map}: For this metric, we used
the following methodology: we considered $p=20$ points closest to a
straight line that goes from the origin of the embedding to a random
chosen point in the Euclidean space (in $d$ dimensions) ordered by the
distance to the farthest point. We then check each consecutive pair of
points (TV shows and movies) for their genders. The objective is to check
how smooth is the transition between the genders in the trajectory. That
is, we define the gender gradient by calculating the gender based
similarity $genderBasedSimilarity(A, B)$ defined in \ref{eq:simGenero}
between two consecutive points of the straight line. In the end we use the average of the $genderBasedSimilarity$ values to define the ``smoothness'' of the line.

\begin{figure}[t]
\begin{center}
  \subfigure[\footnotesize Gender gradient in randomly chosen straight lines.]
            {\includegraphics[width=.65\columnwidth]{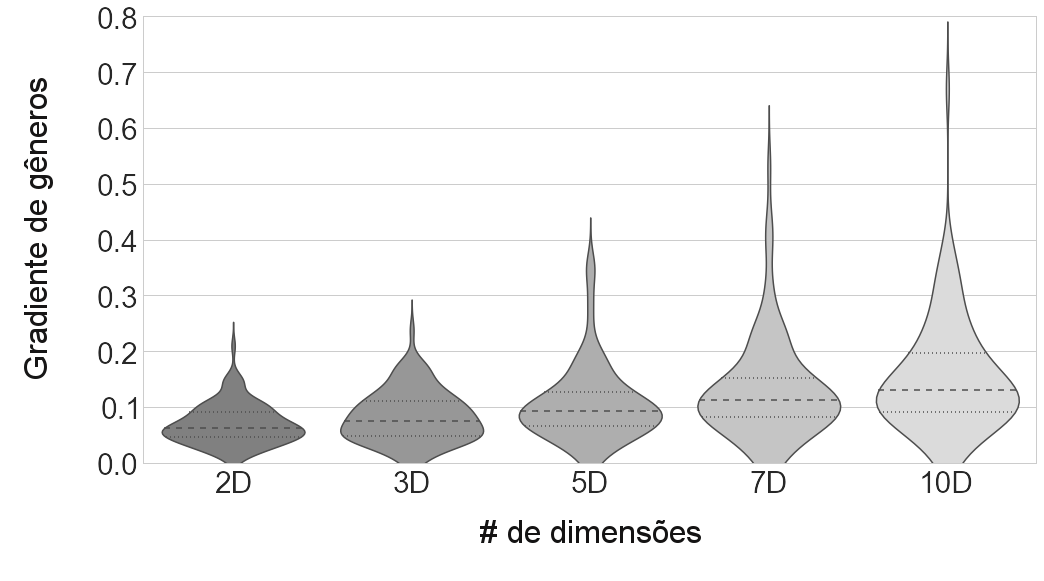}
            \label{fig:ViolinPlot100Lines}}
  \subfigure[\footnotesize Neighborhood gender-based similarity.]
            {\includegraphics[width=.65\columnwidth]{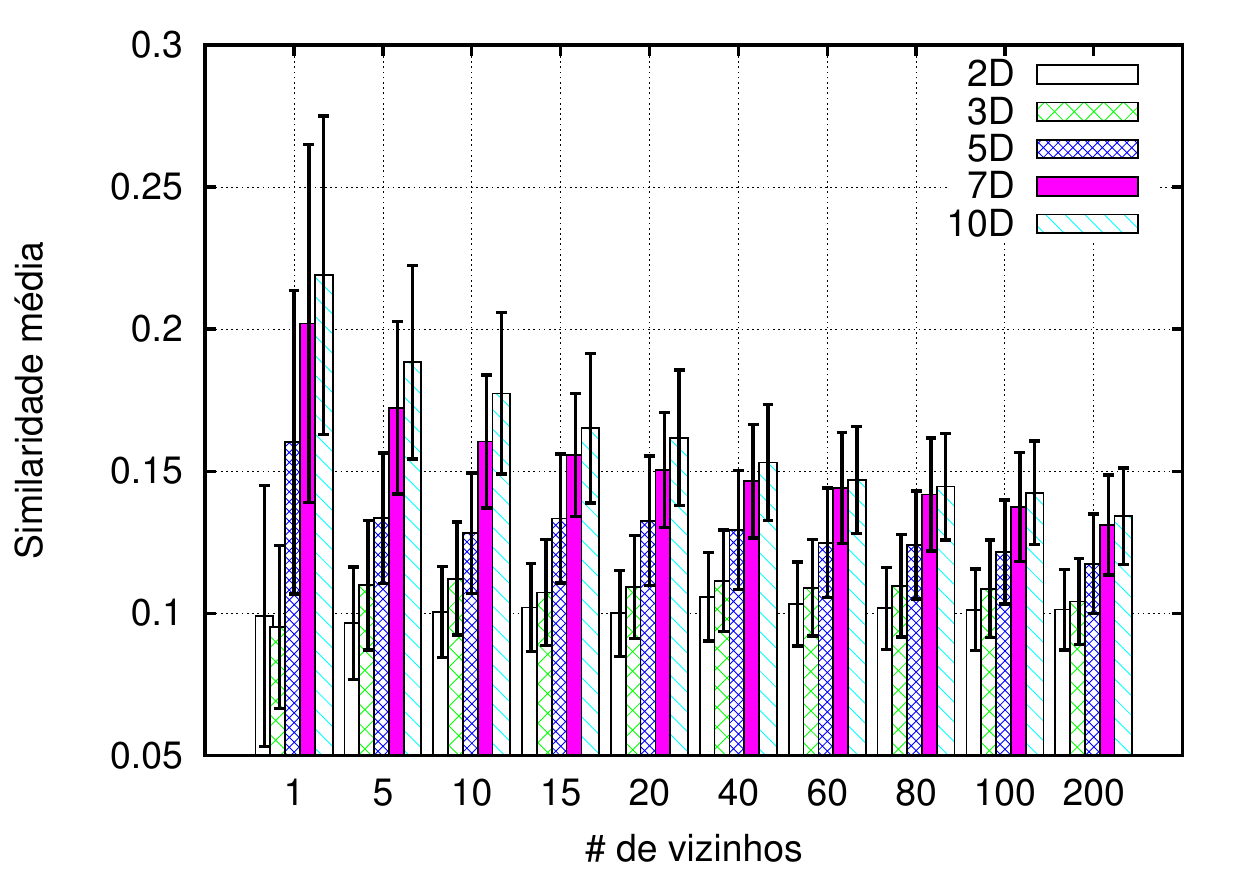}\label{fig:closeGenresSimilarity}}
            \caption{\footnotesize Embedding quality analysis for dimensionality
            }
\end{center}
\end{figure}

Figure~\ref{fig:ViolinPlot100Lines} shows the approximation of the
gender-based similarity distribution for 100 randomly chosen straight
lines for 2, 3, 5, 7 and 10 dimensions. We can see that the gender
gradient is smoother for a higher number of dimensions, indicating that
the genders are more cohesively grouped in maps with higher
dimensionality.

\textbf{Gender predominance in a neighborhood:} We randomly selected 50
items in the embedding and checked the gender distribution in their local
neighborhood, observing different sizes of neighborhood. For each
neighborhood size, we computed the gender-based similarity between the
neighbors and the chosen point, using the definition \ref{eq:simGenero}.
Figure~\ref{fig:closeGenresSimilarity} shows the average gender-based
similarity between the points and their neighbors with 95\% confidence.
We can see that the similarity between a point and its neighbors is
bigger for smaller neighborhoods and higher dimensionality.

%% file: sections/musicDomain.tex
\chapter{Music Domain}
\label{sec:musicDomain}

This chapter will evaluate the maps of media created for the domain of
music based on data collected from Last.fm.

\url{Last.fm} provides a public API for collecting
information about songs, artists, albums and user listening habits.
For this purpose, items are
represented by music tracks and item-to-item similarity reflects how often two
tracks co-occurred in user profiles.

We will analyze how well the constructed Euclidean space represents
music similarity. Note that there are no standard metrics for domain-specific
embedding algorithms. Therefore, we first look into a domain-independent
measure, the residual variance per dimension. Next, we evaluate artist
similarity in local neighborhoods of increasing size.

In the created maps, songs are assigned coordinates and distances express
their (global) similarity.

To analyse the two embedding techniques Isomap and L-Isomap, we will use two
datasets, since the Isomap approach have a high computational cost and can not
handle too much data.

The first dataset contains the users' top 25 most listened songs.
From a total of 2,060,173 songs, 983,010 have MusicBrainz
Identifiers (MBID)\footnote{MBID is a reliable and unambiguous form of music
identification (\url{musicbrainz.org}).}. For the purpose of our study,
we considered a subset of 83,180 tracks that co-occurred 5 or more times,
forming a connected component of 62,352 songs.

The second dataset contains the users' top 100 most listened songs.
From a total of 5,477,927 songs, we considered a subset of 1,520,771
tracks that co-occurred 2 or more times,
forming a connected component of 1,457,786 songs.

The 62k dataset will be used to compare Isomap and L-Isomap, since
we are not able to run the Isomap technique in the 1.4M dataset.

\begin{figure*}[h]
\centering
\includegraphics[width=1\linewidth]{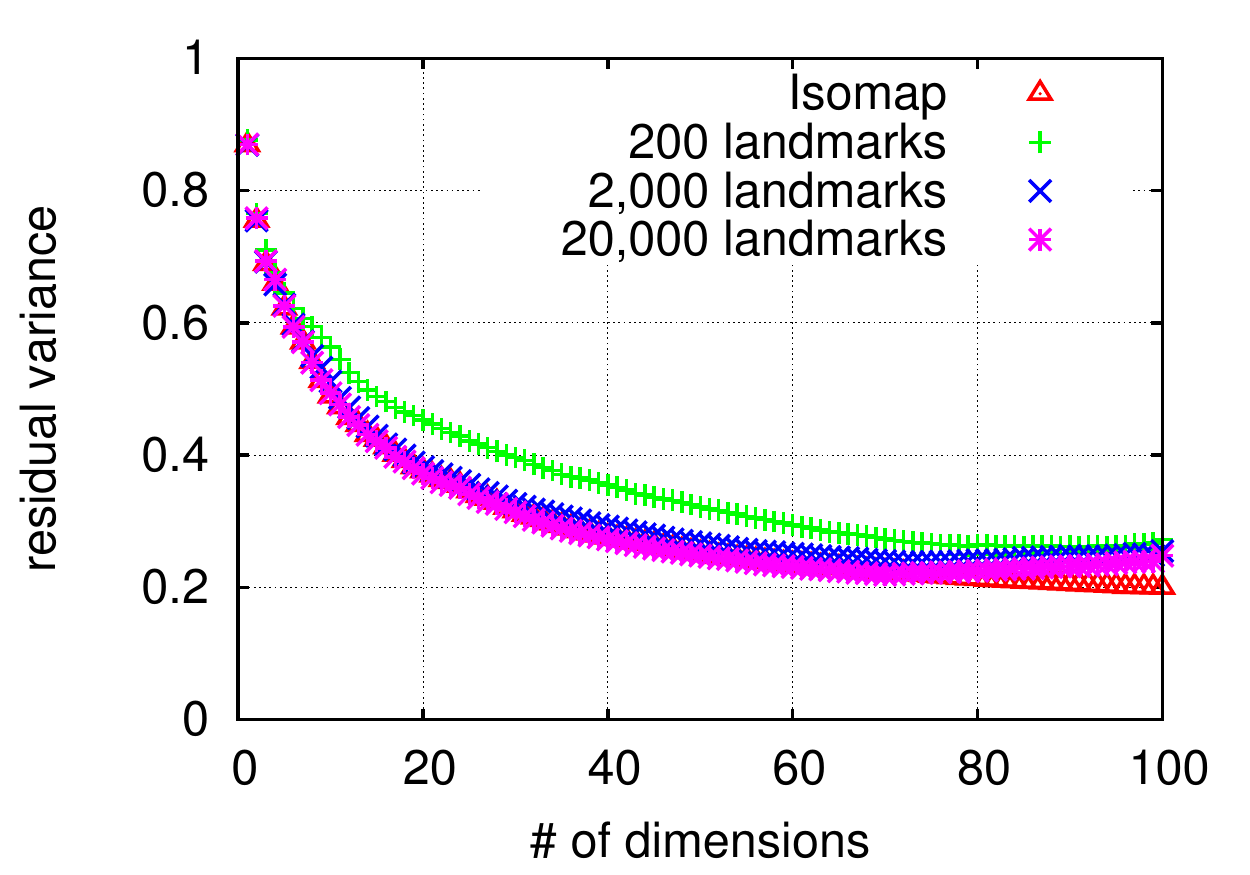}
\caption{Embedding distortion per dimension using Isomap and Random Landmarks L-Isomap with 62k dataset}\label{fig:embeddinganalysisA}
\end{figure*}

Figure~\ref{fig:embeddinganalysisA} depicts the residual
variance~\cite{TenenbaumEtAl2000} associated with the $k$-th dimension
output by the Isomap and L-Isomap algorithm with random landmarks using the 62k dataset, for $k \in [1,\ldots,100]$.
The stronger the decay of variance
from dimension $k-1$ to $k$, the more significant is the addition of the $k$-th
dimension to the map. It can be seen that the residual variance has the
strongest drop in the first 20 dimensions; residual variance up to 20 dimensions tends to 
drop slowly, remaining almost constant for values up to 80 dimensions.

Increasing the number of landmarks from 200 to 20,000 reduces the residual
variance specially for 20 to 60 number of dimensions.  Comparing to the Isomap
residual variance, L-Isomap provides a good
approximation, specially with 20,000 landmarks. We also note that for 20,000
landmarks, the decrease in residual variance seems to reach a plateau around 70
dimensions, and the gain of adding more dimensions to the map is small.

For the 62k dataset we are able to conclude that with only 200 (3\%) random landmarks there
is a good approximation of Isomap.

\begin{figure*}[h]
\centering
\includegraphics[width=1\linewidth]{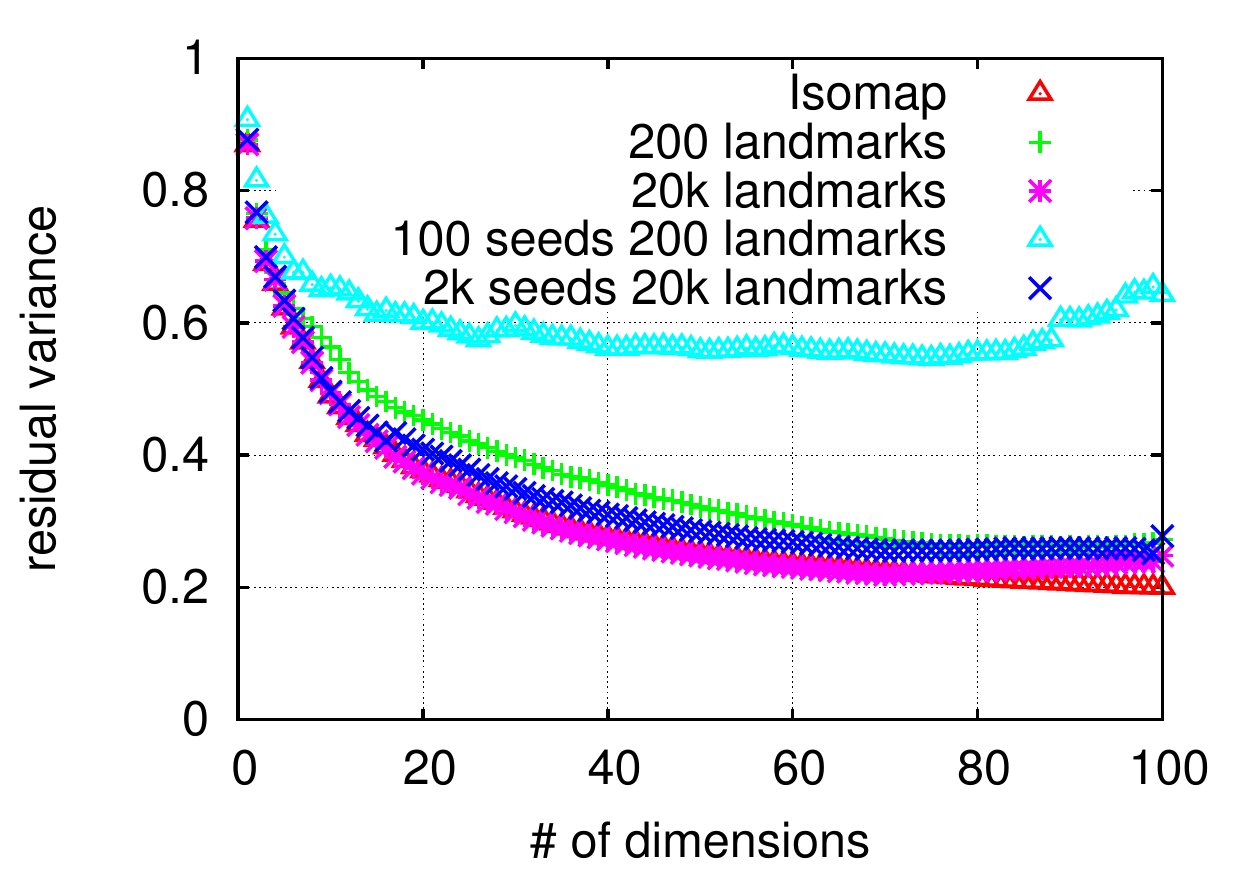}
\caption{Embedding distortion per dimension using Isomap, Random and MaxMin Landmarks L-Isomap with 62k dataset}\label{fig:embeddinganalysisMaxMin}
\end{figure*}

Figure~\ref{fig:embeddinganalysisMaxMin} depicts the residual
variance using the MaxMin function choosing random seeds, Isomap
and full random Landmarks choice.
With only 100 seeds we were
not able to get a good result comparing with the others, but using 2,000
seeds and 20,000 landmarks there is only a small difference between
the full random 20,000 landmarks.
Our hypothesis is that this dataset probably have a uniform distribution
and so the random result is very close to the full Isomap technique and
the MaxMin function. To avaliate better the performance of the 100 seeds
test we need more experiments and analyse a confidence interval.

\begin{figure*}[t]
\begin{minipage}{.48\linewidth}
\centering
\includegraphics[width=1\linewidth]{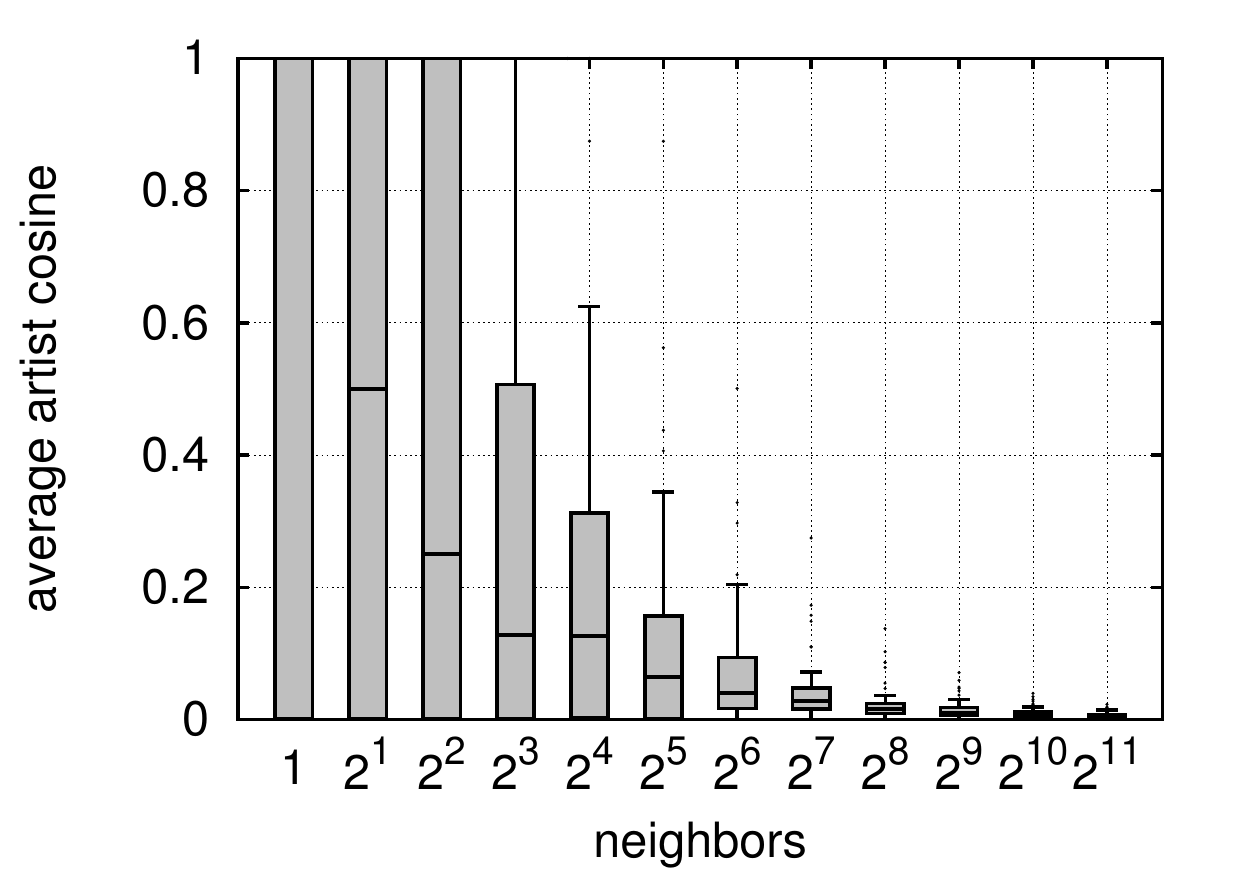}
\caption{Artist similarity per neighborhood size in 10D for the map created with Isomap in 62k dataset}
\label{fig:embeddinganalysisB}
\end{minipage}
\hfill
\begin{minipage}{.48\linewidth}
\centering
\includegraphics[width=1\linewidth]{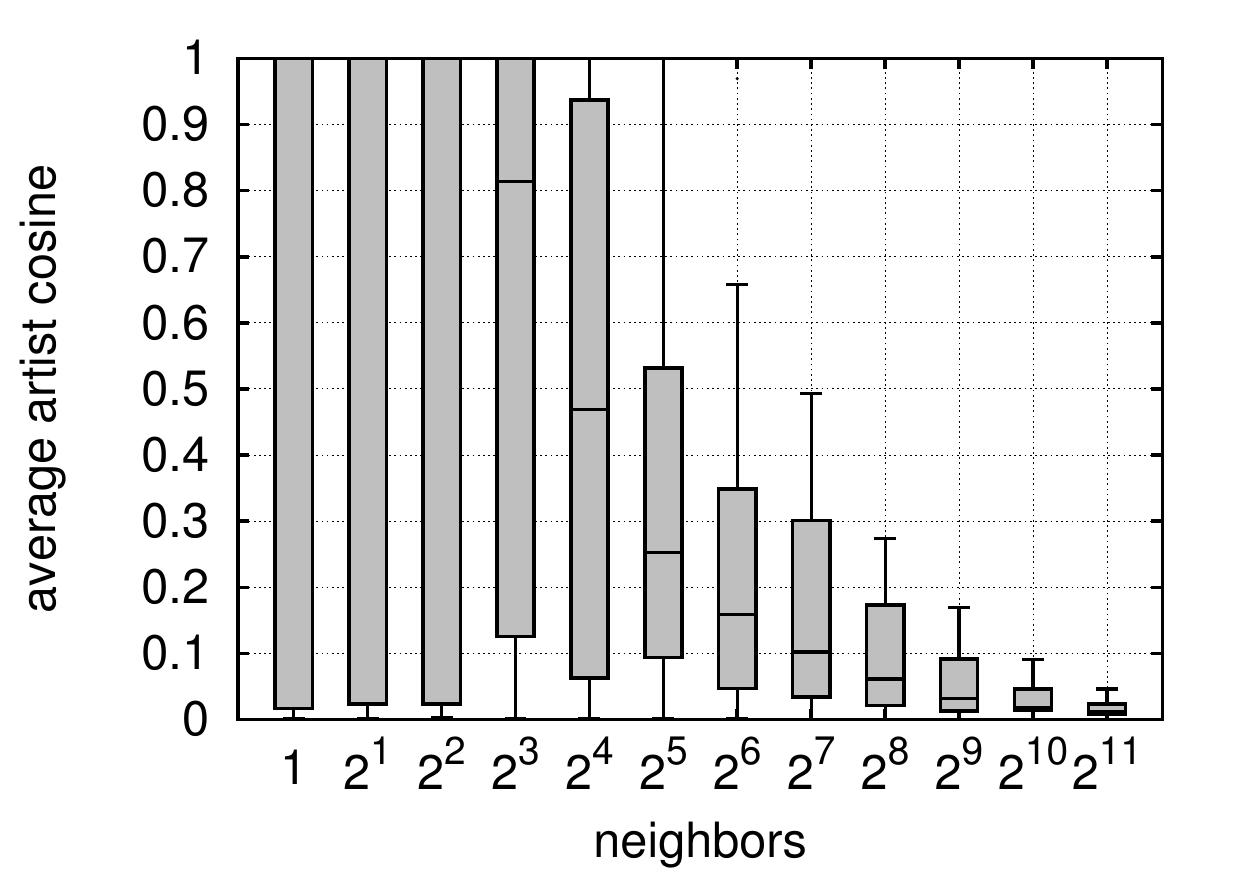}
\caption{Artist similarity per neighborhood size in 100D for the map created with Isomap in 62k dataset}
\label{fig:embeddinganalysisC}
\end{minipage}
\end{figure*}

To further analyze how space dimensionality impacts the item-to-item similarity, 
Figures \ref{fig:embeddinganalysisB} and \ref{fig:embeddinganalysisC} depict a box plot of the average cosine
similarity of artists\footnote{We assume that the more frequently two artists
co-occur in a user’s listening history, the more similar they are. Then, songs from the same artist have similarity equal to 1.} in local
neighborhoods of increasing size, for the 10-dimensional and 100-dimensional spaces, respectively. 
It can be seen that the median similarity of artists in neighborhoods of size 1 and 2 is 1,
i.e., pairs and triples of closest songs tend to belong to the same artist, and
the artist similarity gradually decreases as the size of the neighborhood
increases, in both considered dimensions. This demonstrates that songs composed by the same and similar artists
tend to be clustered on the map, whereas distant nodes tend to belong
to dissimilar artists.

We show that the 100-dimensional space embedding has higher-quality similarity representation.
However, less dimensions can be used if the map is used in
resource-constraint environment, such as mobile devices \cite{kuhn2010social}.

\begin{figure*}[t]
\begin{minipage}{.48\linewidth}
\centering
\includegraphics[width=1\linewidth]{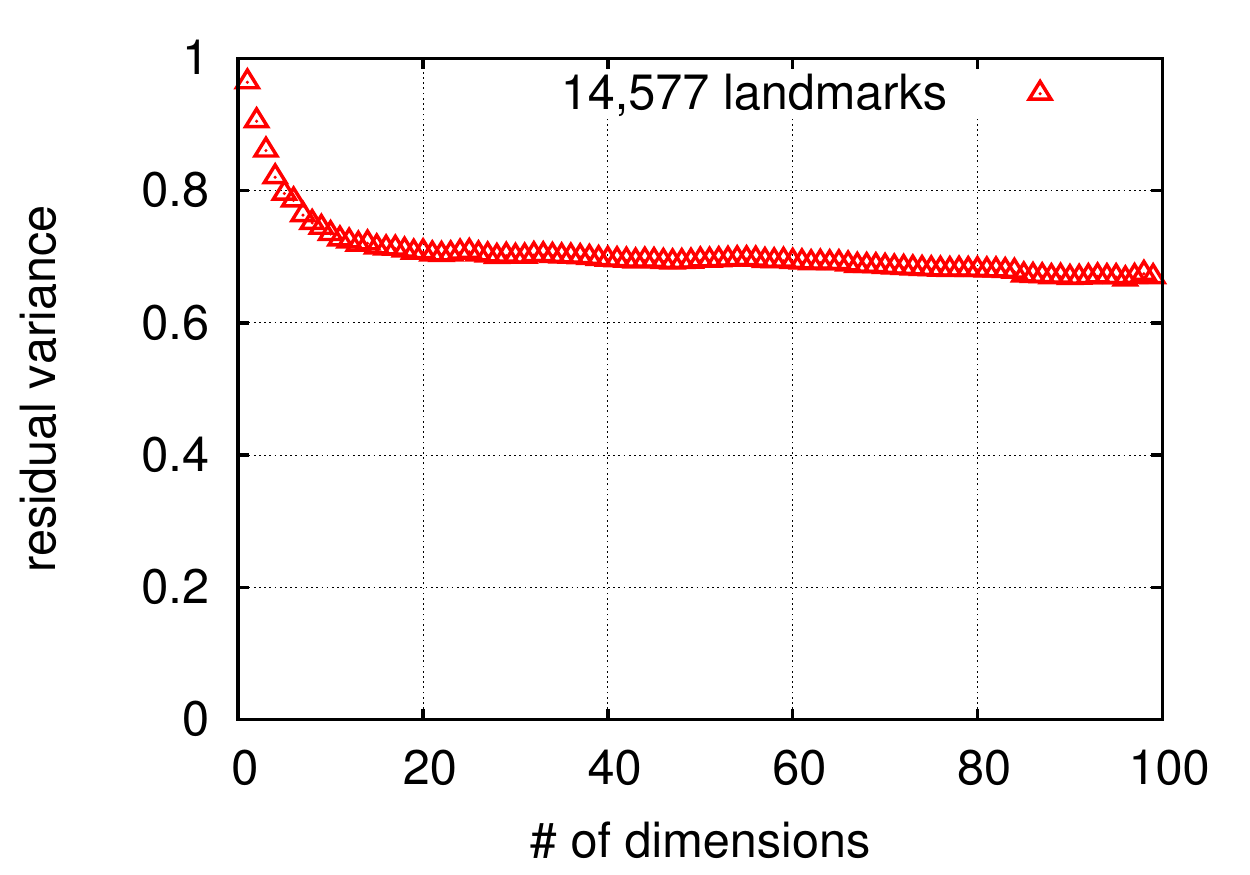}
\caption{Embedding distortion per dimension using Random Landmarks L-Isomap in 1.4M dataset}
\label{fig:embeddinganalysis1M}
\end{minipage}
\hfill
\begin{minipage}{.48\linewidth}
\centering
\includegraphics[width=1\linewidth]{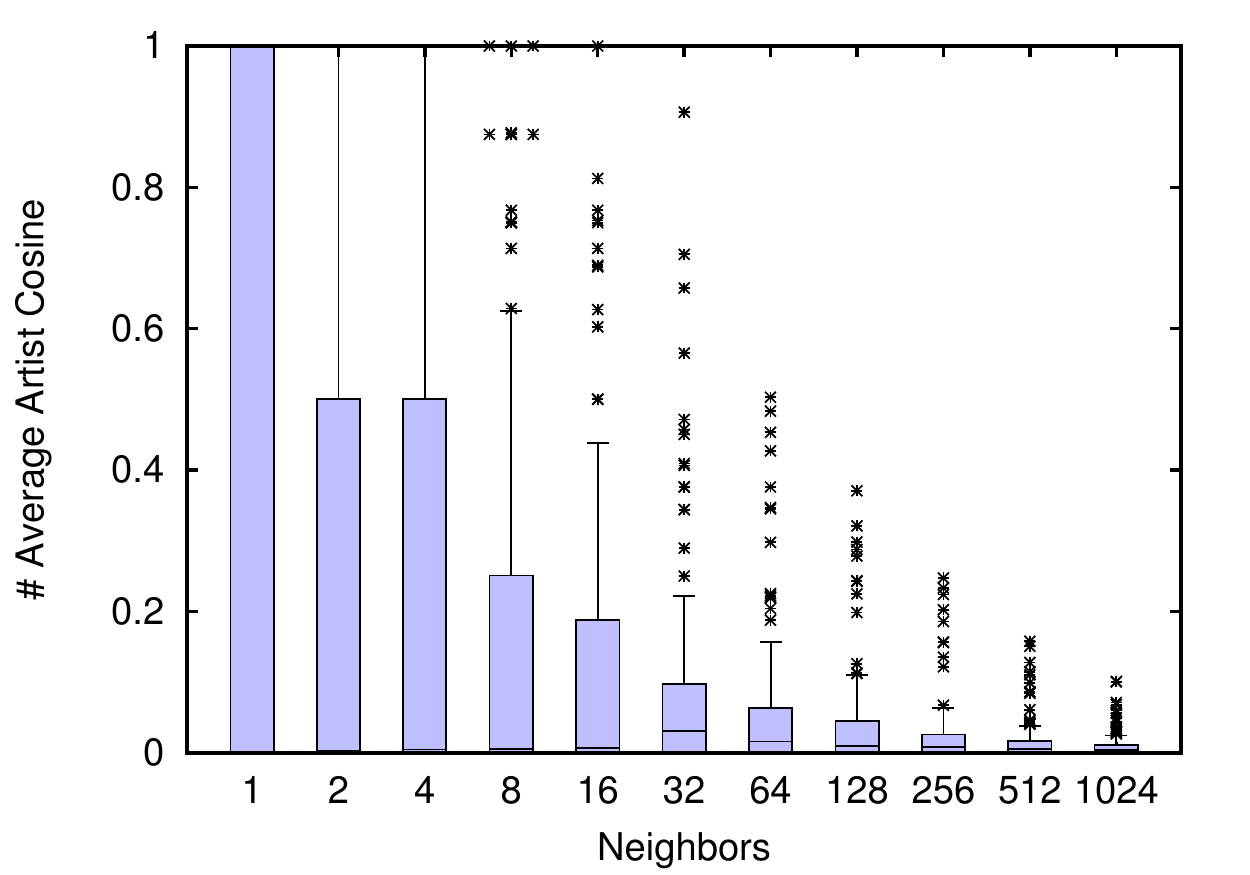}
\caption{Artist similarity per neighborhood size in 100D for the map created with Isomap in 1.4M dataset}
\label{fig:embeddinganalysisNeighbor1M}
\end{minipage}
\end{figure*}

To analyse the L-Isomap performance with more data,
Figure~\ref{fig:embeddinganalysis1M} shows the residual
variance using the random landmarks L-Isomap in the 1.4M
dataset. Using 14,577 landmarks, 1\% of the data, we hit
a minimum of ~66\% of residual variance. Because L-Isomap
is based on a uniform sampling assumption of the data \cite{silva2002global}
we believe that the dataset have `noise' points.

Figure~\ref{fig:embeddinganalysisNeighbor1M} confirm our hypotesis, since
the close neighbors are dissimilar in the average, but
considering 32 or more neighbors it starts to get higher
similarity.

%% file: sections/conclusion.tex
\chapter{Conclusion}\label{sec:conclusion}

In this work we analyzed different nonlinear dimensionality reduction methods applied
on the construction of media similarity maps showing that L-Isomap have a small decrease
in accuracy but with a gain in time complexity.
We intend to further investigate the 1.4M dataset distribution
to remove the `noisy' data and improve the music map quality.

The proposed techniques can be used in different domains, allowing the construction of similarity maps for different purposes.
Also the number of dimensions have direct influence
in the quality of the map, but after adding some dimensions
the decrease in residual variance seems to reach a plateau
and probably this is the intrinsic dimensionality
of the data.
The proposed structure allows:

\begin{itemize}
  \item \textit{Scalability gains:} To store all-pairs items similarity, one
  (multi-dimensional) coordinate needs to be stored for each item, therefore
  resulting in linear ($O(n)$) space complexity. Moreover, memory complexity is $O(k)$ for small subsets of $k$
  items, possibly stored on mobile devices, needing no global information to
  compute similarity values. This is in contrast to quadratic space
  requirements of other data structures, like graphs, which need to be all
  in memory to compute similarity in small subsets of items.
  \item \textit{Performance gains:} With this structure the similarity between two items
  can be calculated in constant time $O(1)$.
  \item \textit{Improved media navigation:} Knowing that similar items are close to
  each other we are able to implement new ways of navigation in a media collection.

\end{itemize}